\documentclass[amssymb,amsmath,showpacs,aps,prl,twocolumn,superscriptaddress]{revtex4}
\usepackage{graphicx}
\usepackage{dcolumn}
\usepackage{bm}
\usepackage[up]{subfigure}

\begin{document}


\title{
Magnetic interactions of supported magnetic clusters
}

\author{Anders Bergman}
\email{Anders.Bergman@fysik.uu.se}
\affiliation{Department of Physics, Uppsala University, Box 530 Sweden}
\author{Lars Nordstr\"om}
\affiliation{Department of Physics, Uppsala University, Box 530 Sweden}
\author{Angela Burlamaqui Klautau}
\affiliation{Departamento de F\'\i sica, Universidade
Federal do Par\'a, Bel\'em,
 PA, Brazil}
 \author{Sonia Frota-Pess\^oa}
\affiliation{Instituto de Fisica, Universidade de Sao Paulo, CP 66318, Sao Paulo, SP, Brazil}
\author{Olle Eriksson}
\affiliation{Department of Physics, Uppsala University, Box 530 Sweden}
\date{\today}

\begin{abstract}
It is demonstrated that the magnetic interactions can be drastically different for nano-sized systems compared to those of bulk or surfaces. Using a real-space formalism we have developed a method to calculate non-collinear magnetization structures and hence exchange interactions. Our results for magnetic clusters supported on a Cu(111) surface show that the magnetic ordering as a rule is non-collinear and can not always be described using a simple Heisenberg Hamiltonian. We suggest that {\it ab initio} calculations allowing for non-collinear coupling between atomic spins is the best tool for analyzing nano-sized magnets.
\end{abstract}
\maketitle
%
The effort of shrinking materials and devices to nano-sizes is fueled both by scientific curiosity and industrial requirements. Applications are found in most scientific fields (photonics and electronics\cite{gudiksen}, biotechnology\cite{berry}, information technology\cite{gambardella}, materials science\cite{barsoum}, and energy applications\cite{poizot})  and devices based on nano-technology are rapidly becoming a natural part of our daily life (e.g. in personal computers).
The best way to characterize a nano-material is, apart from its size reaching nano-meter dimensions, that finite size or quantum effects dominate, yielding new interactions and novel functionality.
\par
Small clusters supported on a surface are of special interest since they have the potential of increasing the density in information storage. One may envision that future magnetic hard discs with information carried by magnetic clusters, will have a storage density two orders of magnitude larger than those used today. The properties of such systems may be measured by means of scanning tunneling microscopy (STM)\cite{binnig}, where information is acquired on an atomic scale and atoms are imaged directly. This technique represents an enormous experimental development, and it has been applied to several nano-magnets,\cite{bode} but it must be followed by complimentary theoretical methods. The complication lies in that, due to the nano-size of these systems, traditional theoretical models based on bulk magnetism are inappropriate. This calls for a first principles method adapted for supported clusters where the constraint to fix the spin arrangement in a collinear way must be released so that complex non-collinear magnetic structures can be analyzed. Such a technique is demonstrated here and it is used for clusters supported on a Cu (111) surface. We have studied a large body of Cr and Mn clusters with different geometries. For practical reasons we present here only the results for the Mn clusters, but our conclusions are general and applicable for any magnetic cluster.
\par
In order to correctly describe the physics of isolated clusters supported on surfaces in an efficient way, the theoretical method should preferably be real-space (RS) based, or at least not depend on translational symmetry. While many methods can treat free clusters, the only method capable of treating supported clusters\cite{stoeffler,robles}, reported so far in the literature, does not treat the non-collinearity in a fully self-consistent way, but relies on external parameters obtained from experiments or other electronic structure calculations. 
The self-consistent non-collinear real space method used in this study is based on the Haydock recursion method\cite{haydock} and the LMTO-ASA technique\cite{andersen}. Besides being formulated completely in real-space, our method also has the advantageous feature of having the computational cost scaling linearly with the number of inequivalent atoms in the system, and can thus be used for calculations of large systems. The RS-LMTO-ASA, a collinear version of this method,  has successfully been used extensively for various types of electronic structure calculations earlier, and a more elaborate description of the collinear implementation can be found in Refs. \onlinecite{sonia1} and  \onlinecite{sonia2}.
\par
The recursion method is, in its original formulation, designed to calculate the local density of states (LDOS), which corresponds to the diagonal terms of the imaginary part of the local Green's function $\mathcal{G}(\epsilon)=(\epsilon-\mathcal{H})^{-1}$ for specified atoms and orbitals, where $\mathcal{H}$ is the Hamiltonian and $\epsilon$ is the energy. For a correct treatment of non-collinear magnetism, evaluation of off-diagonal parts of the Green's function are needed, which in principle are possible to extract using recursion methods. This is quite cumbersome and can be circumvented by using a unitary transformation $\mathcal{U}$ on the Hamiltonian $\mathcal{H}$, $\mathcal{H}'=\mathcal{UHU}^\dagger$. The Green's function is transformed in the same way; $\mathcal{G}'=\mathcal{UGU}^\dagger$. Using the relation $\mathcal{U}^\dagger\mathcal{U}=1$ and the fact that cyclic permutations of matrix multiplications conserve the trace of the product, the magnetic density of states $\bm{m}(\epsilon)$, can be written as 
\begin{equation}
\label{eqn1}
\bm{m}(\epsilon)=-\frac{1}{\pi}\Im tr\{\bm{\sigma}\mathcal{U}^\dagger\mathcal{U}\mathcal{G}\mathcal{U}^\dagger\mathcal{U}\}
=-\frac{1}{\pi}\Im tr\{\bm{\sigma}'\mathcal{G}'\},
\end{equation}
where $\bm{\sigma}$ are the Pauli matrices, $(\sigma_x,\sigma_y,\sigma_z)$ and $\bm{\sigma}'$ is a Pauli matrix after the unitary transformation. The transformation matrix $\mathcal{U}$ is different for the tree directions, and chosen so that, $\mathcal{U}\sigma_j\mathcal{U}^\dagger=\sigma'_z$, for $j=x,y,z$, to yield a diagonal representation. The unitary transformation corresponds to a spin rotation where $\mathcal{U}$ can be calculated using spin-$\frac{1}{2}$ rotation matrices. These transformations can then be applied to the original Hamiltonian for calculating the LDOS along the three orthogonal directions. With the Hamiltonian decomposed in a spin-dependent part, $\mathbf{B}$, and a spin-independent component, $H$, $\mathcal{U}$ operates only on the spin-dependent part,
\begin{equation}
\mathcal{H}' = H  + \bm{B} \cdot \mathcal{U}\bm{\sigma}\mathcal{U}^\dagger.
\label{eqn2}
\end{equation}
From the transformed Hamiltonians, $\mathcal{H}'$, the LDOS for the different directions are obtained using the recursion method, the local magnetization axis is calculated and the LDOS for the local spin axis is constructed. Since our Hamiltonians are constructed within an {\it ab initio} LMTO-ASA formalism, all calculations are fully self-consistent, and the spin densities are treated within the local spin density approximation (LSDA)\cite{vonbarth}. No external parameters are thus needed to perform the calculations within our scheme.  Compared with the collinear case, the computational cost is tripled since the recursion is now performed for three directions, but it stills scales linearly with the number of atoms.  
\par
In this study we have considered a large number of clusters with different shapes and sizes supported on a Cu(111) surface. The calculations of the Mn clusters have been performed by embedding the clusters as a perturbation in a previously self-consistently converged 'clean' Cu(111) surface. The cluster atoms and neighboring Cu atoms are then recalculated self-consistently while the electronic structure for atoms far from the cluster are kept unchanged. As is usually the case for LMTO-ASA methods, the vacuum outside the surface needs to be simulated by having a number of layers of empty spheres above the Cu surface in order to provide a basis for the wave-function in the vacuum and to treat charge transfers correctly. Structural relaxations have not been included in this study, so surface and cluster sites have been placed on a regular fcc lattice with the experimental lattice parameter of Cu. The clean Cu(111) surface has been modeled by a large ($>$5000) slab of atoms and the continued fraction, that occurs in the recursion method, have been terminated with the Beer-Pettifor\cite{beer} terminator after 20 recursion levels.
\par
\begin{figure}[h]
\begin{center}
\subfigure[]{
\includegraphics*[width=0.15\textwidth]{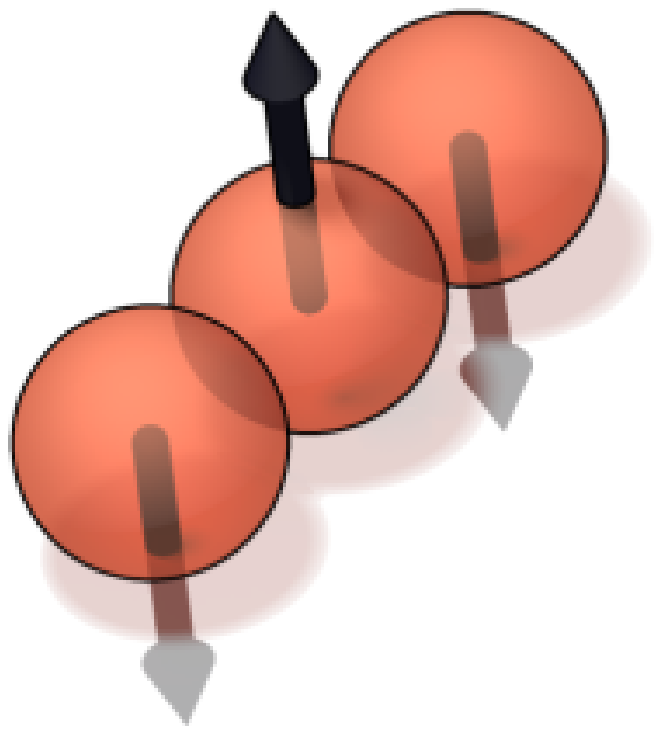}
\label{fig1:a}}
\subfigure[]{
\includegraphics*[width=0.15\textwidth]{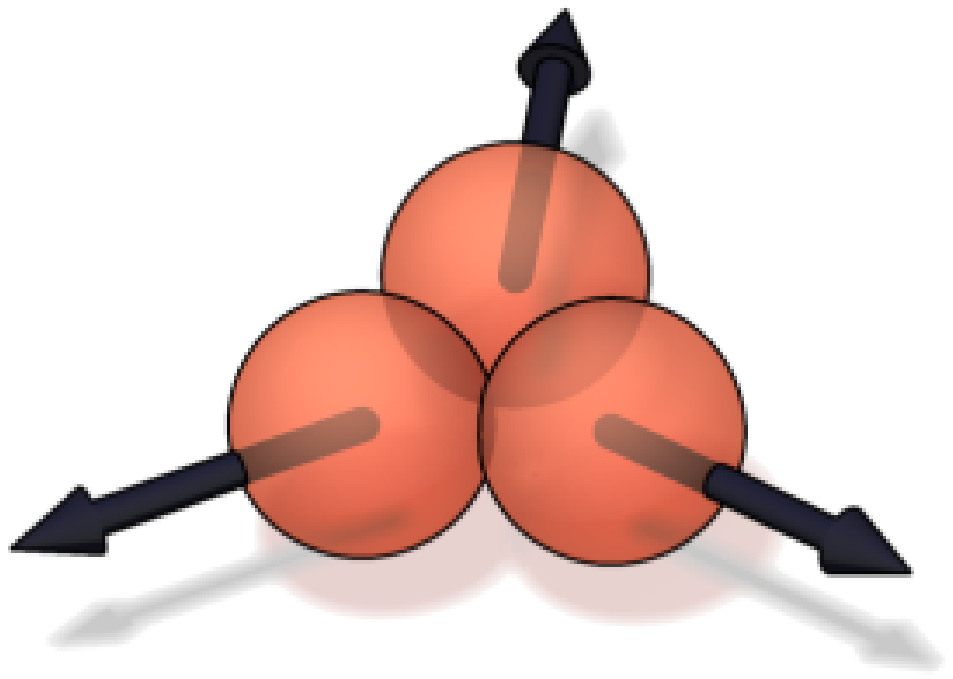}
\label{fig1:b}}\\
\subfigure[]{
\includegraphics*[width=0.15\textwidth]{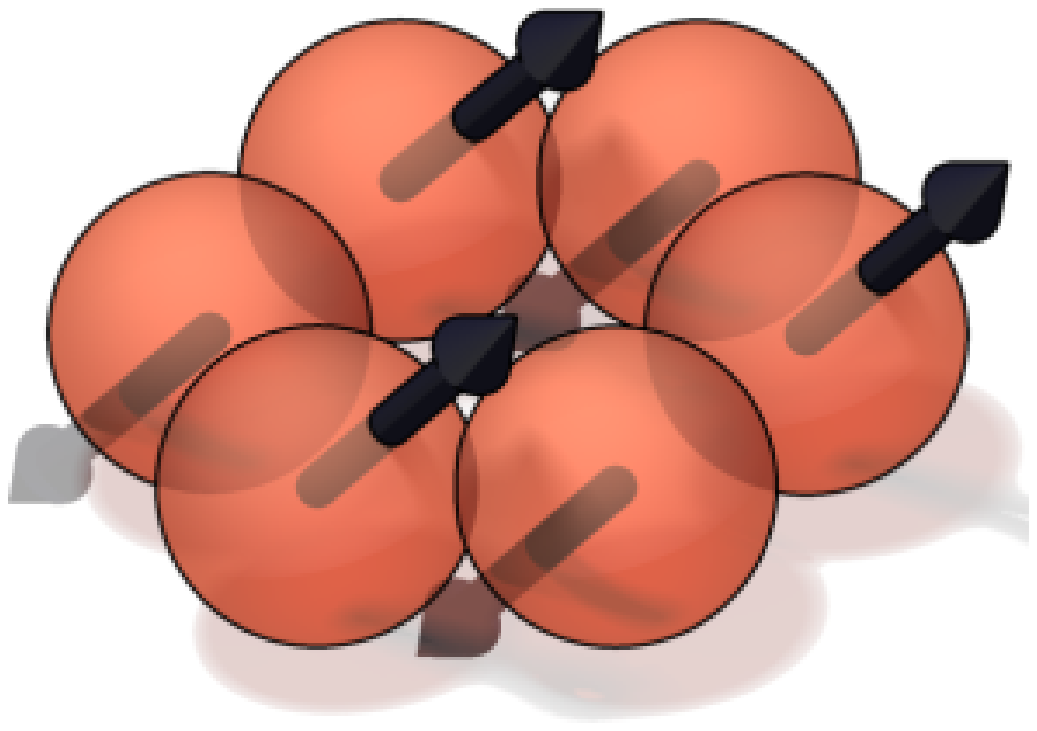}
\label{fig1:c}}
\subfigure[]{
\includegraphics*[width=0.15\textwidth]{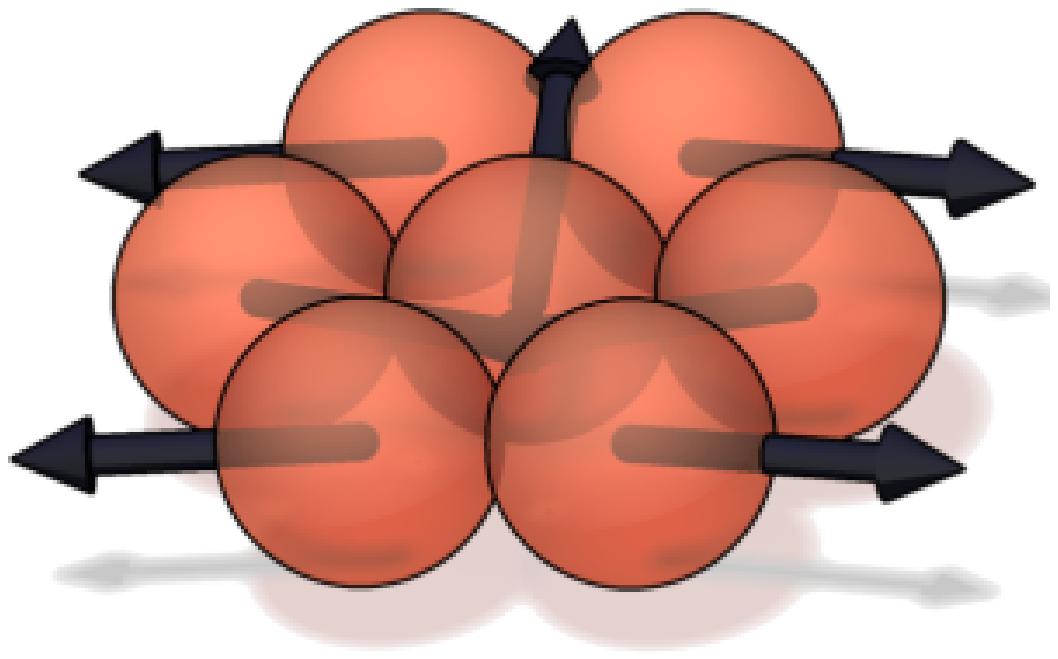}
\label{fig1:d}}\\
\subfigure[]{
\includegraphics*[width=0.2\textwidth]{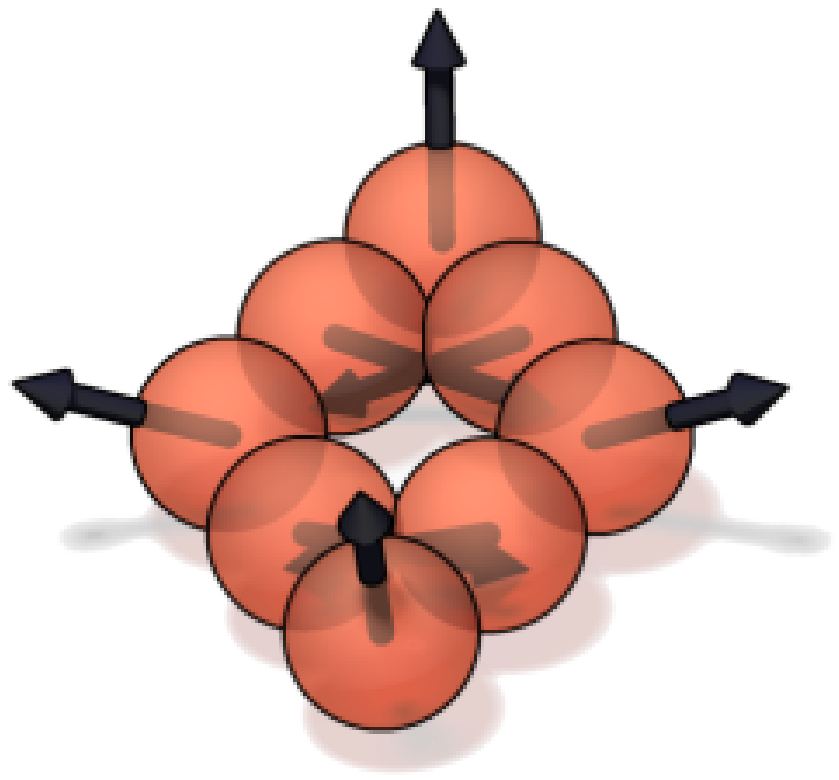}
\label{fig1:e}}
\subfigure[]{
\includegraphics*[width=0.2\textwidth]{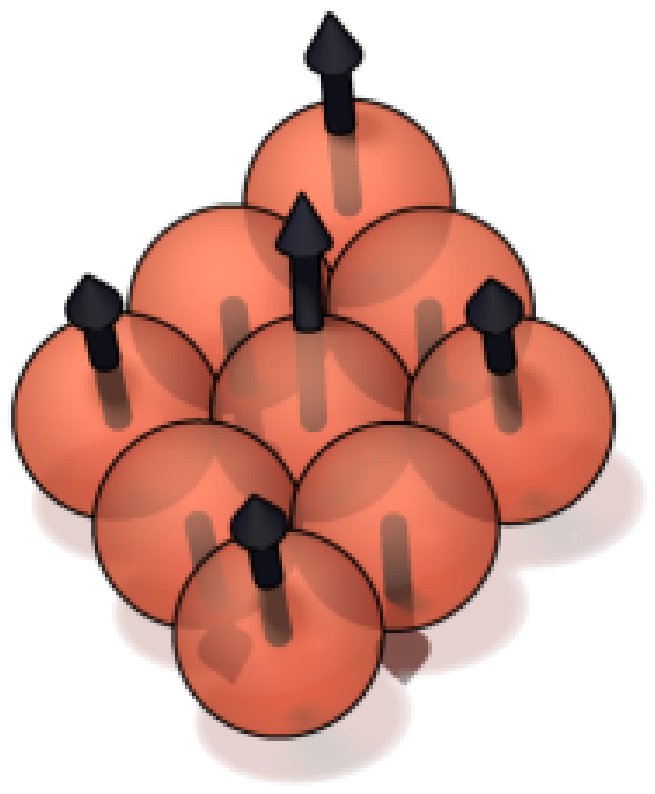}
\label{fig1:f}}\\
\caption{\label{fig:fig1} The magnetic ordering for  Mn clusters on a Cu(111) surface.}
\end{center}
\end{figure}

In Fig.~\ref{fig:fig1} we show Mn clusters with a particularly complex magnetic structure. The local magnetic moments for the Mn atoms in the clusters depend strongly on the number of Mn neighbors with values between $4.7 \mu_B$ for the single adatom to  $2.7 \mu_B$ when six Mn neighbors are present. In Fig.~\ref{fig1:a} the magnetic moments of a linear chain of Mn atoms is shown. Each Mn atom couples its magnetic moment antiparallel to its neighbor in a collinear way. This is in accordance with a Heisenberg Hamiltonian with antiferromagnetic nearest neighbor interactions.
In Fig.~\ref{fig1:b} the three atoms have been moved to form a triangular geometry and the magnetization profile then becomes a non-collinear structure. This is the result of a well-known phenomenon; magnetic frustration\cite{bramwel}. 
In a triangular geometry two magnetic moments with antiferromagnetic interactions can couple antiparallel, but the third moment can not simultaneously be antiparallel to the first two, it becomes 'frustrated'. Instead each moment forms an equal angle of 120 degrees to its neighbor, and the calculated magnetization profile in Fig.~\ref{fig1:b} (and 1a) is the consequence of antiferromagnetic Mn-Mn interactions. 
\par
In Figs.~\ref{fig1:c} and \ref{fig1:d} a more interesting scenario is found. First six Mn atoms forming a hexagonal ring structure were studied (1c). The antiferromagnetic nearest neighbor interaction cause a magnetic order where every second atom has its magnetic moment pointing up and every other has a moment pointing down, and the magnetic order is collinear. However, for the cluster with one extra atom in the center of the hexagonal ring a different magnetic order, with a non-collinear  component, is found. The atoms at the edge of the cluster have a canted anti-ferromagnetic profile, with a net moment pointing antiparallel to the magnetisation direction of the atom in the center of the cluster. The magnetic moment of the central atom is almost perpendicular ($\sim$ 100 $^{\circ}$) to the atoms at the edge of the cluster and with a magnetic moment of 2.7~$\mu_B$. The edge atoms have a magnetic moment of 4~$\mu_B$ per atom that has an angle of $\sim$ 165 $^{\circ}$ to neighboring edge atoms and is parallel to its second nearest neighbors.
\par
In Fig.~\ref{fig1:e}, the magnetic order of a cluster with rhombic shape is shown.  A non-collinear magnetic structure, as is shown in the figure, is found to be metastable in our calculations, and a collinear antiferromagnetic solution has a somewhat lower energy of $\sim 40$ meV/cluster. In Fig.~\ref{fig1:f} we show the moment profile of the rhombic cluster with one extra atom in the middle of the cluster. The most stable magnetic configuration for this cluster is non-collinear with a slightly canted antiferromagnetic structure. 
\par
\begin{figure}
\begin{center}
{
\includegraphics*[width=0.4\textwidth]{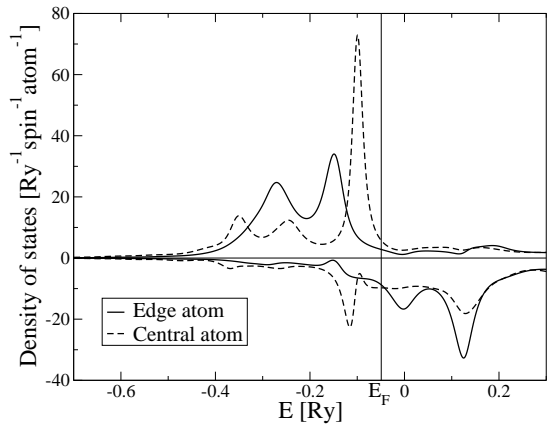}
}
\caption{\label{fig:dos}Local density of states for a central and an edge atom in the cluster in Fig. 1d.} 
\end{center}
\end{figure}

Taken together, the results in Fig.~\ref{fig:fig1} show that as a rule non-collinear ordering is obtained, in contrast to bulk and surface magnets. Also, Fig.~\ref{fig:fig1} shows that an analysis of the magnetism of supported clusters based upon conventional theoretical models, such as the Heisenberg model, is difficult. Due to differences in the local symmetry the electronic structure at different atoms in the cluster becomes unique and different from the neighboring atoms, which can be seen in Fig.~\ref{fig:dos} where the LDOS for the central atom and an edge atom of the cluster in Fig.~\ref{fig1:d} is shown. This means that it is inappropriate to use the same magnetic moment and exchange parameters of e.g. a Heisenberg Hamiltonian for the analysis of the magnetism of the cluster in Fig.~\ref{fig1:d}. Instead, to describe the magnetic properties accurately it becomes necessary to take the step to a non-collinear first principles theory, as done here.
\par
\begin{figure}
\begin{center}
{
\includegraphics*[width=0.2\textwidth]{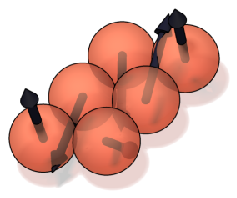}
\includegraphics*[width=0.2\textwidth]{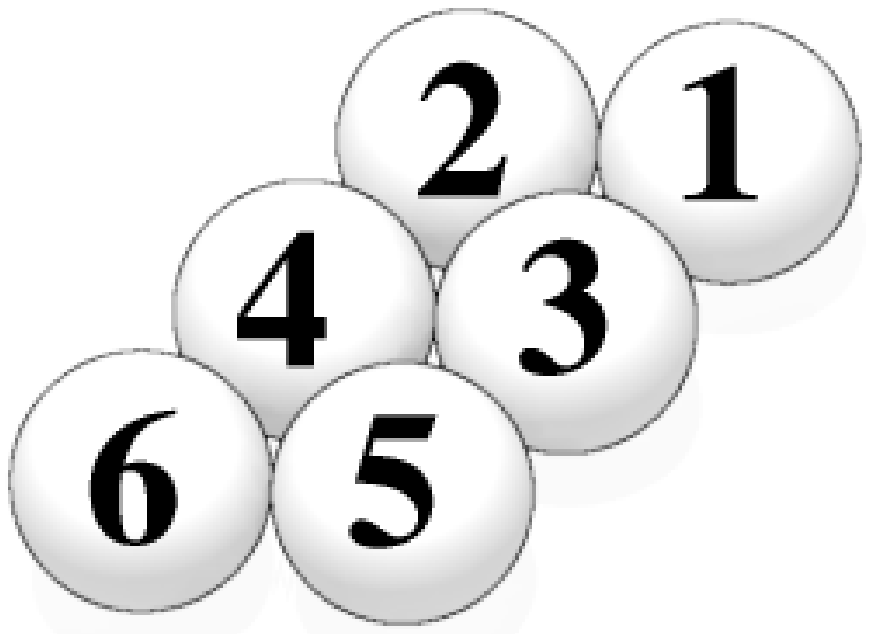}
}
\caption{\label{fig:rodF} Magnetic moments for two triangular Mn clusters connected to each other, resulting in a single, six atom, cluster. The angles between the magnetic moments are given in Table~\ref{table:rodT}, where the atoms are labeled according to the figure to the right.}
\end{center}
\end{figure}

In our calculations, both the exchange interactions within each cluster as well as between the clusters appear naturally. This has been used for examining the distance dependence of inter-cluster interactions of triangular Mn clusters supported on a Cu(111) surface. In Fig.~\ref{fig:rodF}, the two triangular clusters are connected, forming a single, six atom, cluster. As expected, this single cluster shows a non-collinear order due to its frustrated geometry. 
\begin{table}
\caption{\label{table:rodT} Magnetic moments (in $\mu_B$) and angles between moments for the cluster displayed in Fig.~\ref{fig:rodF}. }
\begin{center}
\begin{tabular}{cccccccc}
 Atom & 1 & 2 & 3 & 4 & 5 & 6 & Moment\\
\hline
1 &   0 & 155 &  88 &  80 & 155 &   5 & 4.28 \\
2 & 155 &   0 & 118 &  75 &  50 & 153 & 3.94 \\
3 &  88 & 118 &   0 & 167 &  68 &  89 & 3.60 \\
4 &  80 &  75 & 167 &   0 & 124 &  79 & 3.60 \\
5 & 155 &  50 &  68 & 124 &   0 & 156 & 3.94 \\
6 &   5 & 153 &  89 &  79 & 156 &   0 & 4.28 \\
\end{tabular}
\end{center}
\end{table}

\begin{figure}
\begin{center}
\subfigure[]
{
\includegraphics*[width=0.2\textwidth]{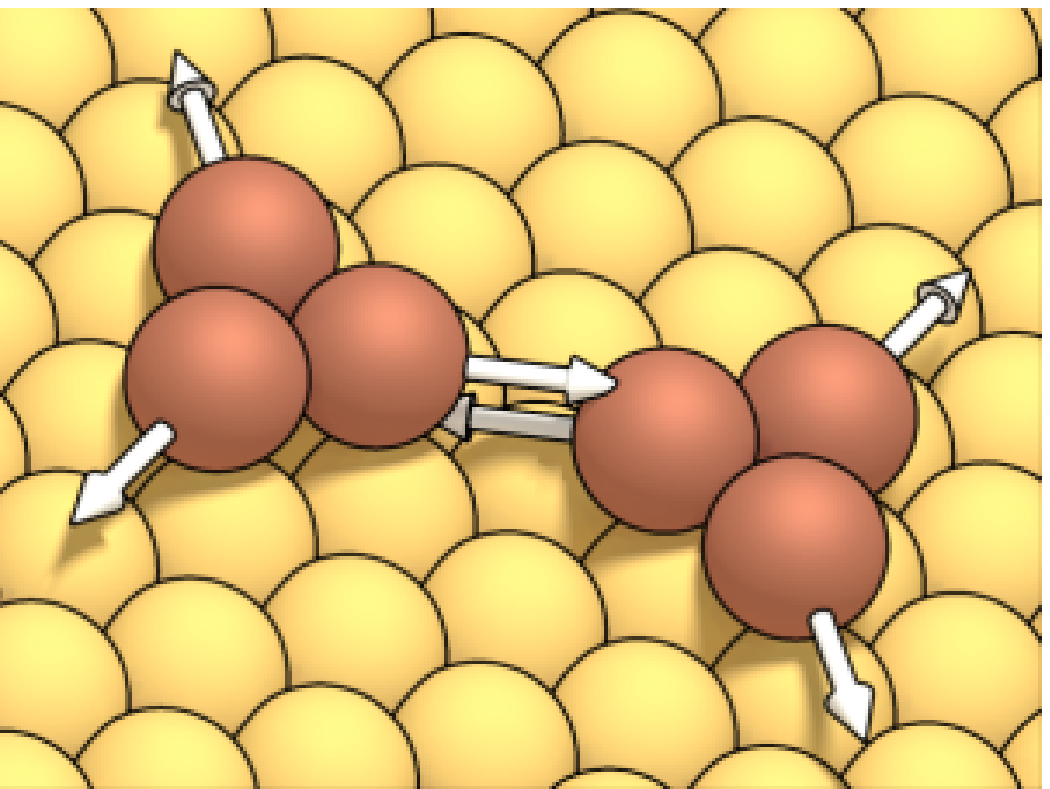}
\label{triF:a}
}
\subfigure[]
{
\includegraphics*[width=0.2\textwidth]{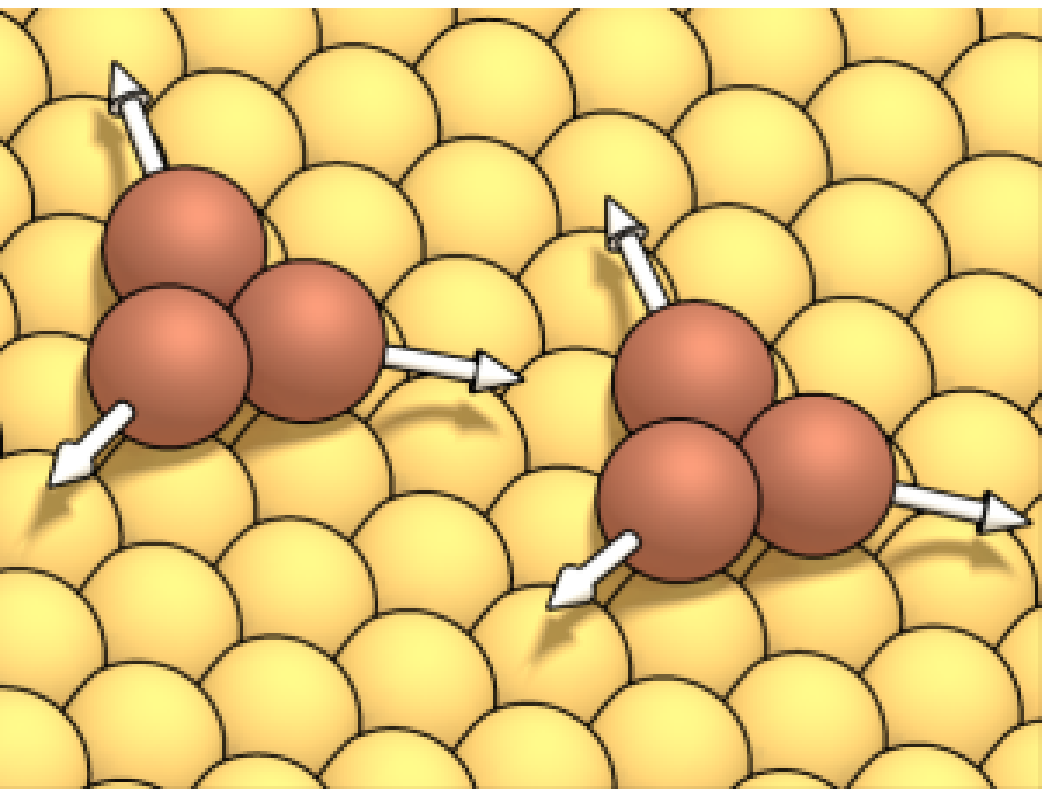}
\label{triF:b}
}
\caption{\label{fig:triF} Magnetic moments for Mn atoms in exchange coupled clusters on a Cu(111) surface. }
\end{center}
\end{figure}

Fig.~\ref{fig:triF} shows the magnetic order of the clusters as they are separated from each other with an increasing distance. 
When the distance between the clusters is very large one expects that the moment profile within each cluster would be identical to that in Fig.~\ref{fig1:b}. The magnetic order of the clusters in Fig.~\ref{triF:a} is indeed close to the non-interacting cluster order with only a few degrees deviation of the magnetic moments. The magnetic moments for the clusters in Fig. 4b are even closer to the magnetic order found for the non-interacting case. It can also be noted that the two cluster-cluster distances in Fig. 4a and 4b give very similar magnetization geometries. This demonstrates that the intra-cluster interaction is much larger than the inter-cluster interaction, even for short cluster-cluster distances. As a measure of the inter-cluster exchange for this geometry, total-energy calculations show that the energy needed to flip the spins by 180 degrees in one of the clusters, in Fig.~\ref{triF:a}, is in the range of a few meV. 
\par
The results in Fig.~\ref{fig:triF} show that the 'frustrated geometry' of the triangular clusters, with 120 degrees between the different moments of each cluster, is very robust, even if the cluster-cluster distance becomes very small. Hence the intra exchange interaction is much stronger (and antiferromagnetic) than the inter exchange interaction.  This finding is important when one attempts to design cluster based media for magnetic information storage where a bit stored in one cluster should not be allowed to interact and degrade the information in the nearest neighboring cluster. 
\par
We end this Letter with a short discussion on the possibility to measure magnetic structures with atomic resolution. 
Spin-polarised STM\cite{SP-STM} has been shown to have the necessary resolution to determine magnetization directions between different atoms in a nanostructure. This technique was also used to measure the g-value and spin excitation at single absorbed Mn atoms on a substrate of Al$_2$O$_3$\cite{heinrich}. Hence STM is the most promising experimental technique to verify the here predicted magnetic profiles.
\par
We acknowledge financial support from the G\"oran Gustafsson foundation, the Swedish Research Council, the Swedish Foundation for Strategic Research, Seagate Technology, Bloomington and  CNPq, Brazil. The calculations were performed at the high performance computing centers UPPMAX, NSC and HPC2N within the Swedish National Infrastructure for Computing and at the computational facilities of the LCCA, University of S\~ao Paulo and of the CENAPAD at University of Campinas, SP, Brazil.

\end{document}